\documentclass{article}
\usepackage{lnfprep}
\usepackage{epsfig}
\usepackage{subfigure}
\def\beq{\begin{equation}}   \def\eeq{\end{equation}}
\def\bea{\begin{eqnarray}}   \def\eea{\end{eqnarray}}

\newcommand{\gsim}{\lower.7ex\hbox{$ \;\stackrel{\textstyle>}{\sim}\;$}}
\newcommand{\lsim}{\lower.7ex\hbox{$ \;\stackrel{\textstyle<}{\sim}\;$}}

\def\c2{CLEO~II.V}
\def\ccb{ c\bar{c} }
\def\d0d0{ D^0\bar{D}^0 }
\def\p0p0{ P^0\bar{P}^0 }

\def\qp2{ \Bigl| \frac{q}{p} \Bigr|^2 }
\def\pq2{ \Bigl| \frac{p}{q} \Bigr|^2 }

\def\ps2s{  \psi(2S) }
\def\q2{ $q^2$ }
\def\cm2s1{ $\,{\rm cm}^{-2} {\rm s}^{-1}$}
\def\d0{D_2^{*0}}
\def\d+{D_2^{*+}}

\def\mev{ \,{\rm MeV}/c^2  }

\newcommand{\krzb}{\ensuremath{\overline{K}{}^*(892)^0 }}
\newcommand{\krzmndk}{\ensuremath{D^+ \rightarrow \krzb \mu^+ \nu}}
\newcommand{\krzlndk}{\ensuremath{D^+ \rightarrow \krzb \ell^+ \nu_\ell}}
\newcommand{\philndk}{\ensuremath{D_s^+ \rightarrow \phi(1020)\; \ell^+ \nu_\ell}}
\newcommand{\phimndk}{\ensuremath{D_s^+ \rightarrow \phi(1020)\; \mu^+ \nu }}
\newcommand{\kkmndk}{\ensuremath{D_s^+ \rightarrow K^+ K^- \mu^+ \nu }}

\newcommand{\gevcsq}{\ensuremath{\textrm{GeV}/c^2}}

\newcommand{\thv}{\ensuremath{\theta_\textrm{v}}}
\newcommand{\thl}{\ensuremath{\theta_\ell}}
\newcommand{\costhv}{\ensuremath{\cos\thv}}

\newcommand{\costhl}{\ensuremath{\cos\thl}}

\newcommand{\qsq}{\ensuremath{q^2}}

\newcommand{\mkk}{\ensuremath{m_{K^+ K^-}}}

\newcommand{\rvvalue}{\ensuremath{1.549 \pm 0.250 \pm 0.145}}
\newcommand{\rtwovalue}{\ensuremath{0.713 \pm 0.202 \pm 0.266}}

\newcommand{\rtwo}{\ensuremath{r_2}}
\newcommand{\rthree}{\ensuremath{r_3}}
\newcommand{\rvee}{\ensuremath{r_v}}

\newcommand{\Header}{
  \begin{tabular}{rl}
  \hspace{-.4cm}
  \includegraphics{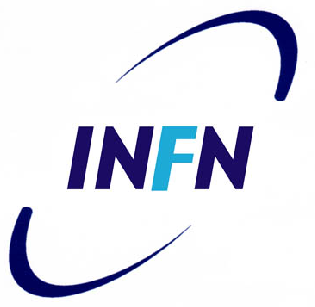} 
      &
    \renewcommand{\arraystretch}{0.5}
    \begin{tabular}{r}
      {\hspace{1cm}~\LARGE\sffamily LABORATORI~ NAZIONALI~ DI~ FRASCATI}\\
      \\
      {\Large\sffamily SIS-Pubblicazioni}\\
    \end{tabular}
    \renewcommand{\arraystretch}{1}
  \end{tabular}
  \vskip 1cm
  \begin{flushright}
  \renewcommand{\arraystretch}{0.5}
    \begin{tabular}{r}
      {\underline{LNF-04/14 (P)}}\\    
      {\small 28 luglio 2004} \\      
      {\tt hep-ex}\\
    \end{tabular}
  \end{flushright}
  \renewcommand{\arraystretch}{1}
  \vskip 1 cm
  }

\begin{document}
\begin{titlepage}
\title{ 
  \Header
  {\large \bf  NEW CHARM RESULTS FROM FOCUS}
}
\author{Stefano Bianco \\
{\em Laboratori Nazionali di Frascati dell'INFN } \\
on behalf$^1$ of the FOCUS Collaboration 
}
\maketitle
\baselineskip=14pt

\begin{abstract}
New results from the 
photoproduction experiment FOCUS are reported: Dalitz
 plot analysis, semileptonic form factor ratios and excited meson spectroscopy.
\end{abstract}

\vspace*{\stretch{2}}
\begin{flushleft}
  \vskip 2cm
{ PACS:} 
\end{flushleft}
\begin{center}
Presented at the 
18th Rencontres de Physique de la Vallee d'Aoste \\
 29 February-6 March, La Thuile, Vallee d'Aoste, Italy.
\end{center}
\end{titlepage}
\pagestyle{plain}
\setcounter{page}2
\baselineskip=17pt
I report\footnote{\scriptsize 
co-authors are: 
J.~M.~Link,  
P.~M.~Yager 
(\textbf{UC Davis});
J.~C.~Anjos, 
I.~Bediaga, 
C.~G\"obel, 
A.~A.~Machado,
J.~Magnin, 
A.~Massafferri, 
J.~M.~de~Miranda, 
I.~M.~Pepe, 
E.~Polycarpo,
A.~C.~dos~Reis 
(\textbf{CBPF, Rio de Janeiro});
S.~Carrillo, 
E.~Casimiro, 
E.~Cuautle
A.~S\'anchez-Hern\'andez, 
C.~Uribe, 
F.~V\'azquez 
(\textbf{CINVESTAV, Mexico City});
L.~Agostino,
L.~Cinquini, 
J.~P.~Cumalat, 
B.~O'Reilly, 
I.~Segoni, 
K.~Stenson 
(\textbf{CU Boulder});
J.~N.~Butler, 
H.~W.~K.~Cheung, 
I.~Gaines, 
P.~H.~Garbincius, 
L.~A.~Garren, 
E.~Gottschalk, 
P.~H.~Kasper, 
A.~E.~Kreymer, 
R.~Kutschke,
M.~Wang 
(\textbf{Fermilab});
L.~Benussi,
M.~Bertani,
S.~Bianco, 
F.~L.~Fabbri, 
A.~Zallo, 
(\textbf{INFN L.~N. Frascati});
M.~Reyes
(\textbf{Guanajuato, Mexico});
C.~Cawlfield, 
D.~Y.~Kim, 
A.~Rahimi, 
J.~Wiss, 
R.~Gardner,
A.~Kryemadhi
(\textbf{UI Champaign});
C.~H.~Chang,
Y.~S.~Chung, 
J.~S.~Kang, 
B.~R.~Ko, 
J.~W.~Kwak, 
K.~B.~Lee
(\textbf{Korea University, Korea});
K.~Cho,
H.~Park
(\textbf{Kyungpook University, Korea});
G.~Alimonti, 
S.~Barberis,
M.~Boschini, 
A.~Cerutti,
P.~D'Angelo, 
M.~DiCorato, 
P.~Dini, 
L.~Edera,
S.~Erba,
M.~Giammarchi, 
P.~Inzani, 
F.~Leveraro, 
S.~Malvezzi, 
D.~Menasce, 
M.~Mezzadri,  
L.~Moroni, 
D.~Pedrini, 
C.~Pontoglio, 
F.~Prelz, 
M.~Rovere, 
S.~Sala, 
(\textbf{INFN and Milano});
T.~F.~Davenport~III, 
(\textbf{UNC Asheville});
V.~Arena, 
G.~Boca, 
G.~Bonomi, 
G.~Gianini, 
G.~Liguori, 
D.~Lopes-Pegna,
M.~M.~Merlo, 
D.~Pantea, 
S.~P.~Ratti, 
C.~Riccardi, 
P.~Vitulo
(\textbf{INFN and Pavia});
H.~Hernandez, 
A.~M.~Lopez, 
H.~Mendez, 
A.~Paris, 
J.~Quinones, 
E.~Ramirez
Y.~Zhang, 
(\textbf{Mayaguez, Puerto Rico});
J.~R.~Wilson, 
(\textbf{USC Columbia});
T.~Handler, 
R.~Mitchell
(\textbf{UT Knoxville});
D.~Engh, 
M.~Hosack, 
W.~E.~Johns, 
E.~Luiggi,
M.~Nehring,
P.~D.~Sheldon, 
E.W.~Vaandering,
M.~Webster, 
(\textbf{Vanderbilt});
M.~Sheaff, 
(\textbf{Wisconsin Madison}).}
 on                                                                                                                             
three new results from the photoproduction experiment FOCUS: the first Dalitz 
plot analysis
 of charm meson decays
 using the K-matrix approach\cite{Link:2003gb}, new
 measurements 
of the \phimndk{} form factor ratios\cite{Link:2004qt},
  and new measurements
 on L=1 excited 
 meson spectroscopy\cite{Link:2003bd}, i.e., 
 precise measurements of the masses and widths of
 the $D_2^{*+}$ and $D_2^{*0}$ mesons, 
 and   evidence for broad states decaying to $D^+\pi^-, D^0\pi^+$ (the  first
   such evidence in $D^0\pi^+$).
    The data for this paper were collected in the Wideband photoproduction
experiment FOCUS during the Fermilab 1996--1997 fixed-target run.
\section{  Dalitz plot analysis of $D_s^+$ and $D^+$ decay to
 $\pi^+\pi^-\pi^+$ using  the \emph{K}-matrix formalism }
Charm-meson decay dynamics has been extensively studied in the
 last decade. The
analysis of the three-body final state by fitting Dalitz plots has proved to be
a powerful tool for investigating effects of resonant substructure,
interference patterns, and final state interactions in the charm sector
\cite{Bianco:2003vb,Malvezzi:2003jp}. The
isobar formalism, which has traditionally been applied to charm amplitude
analyses, represents the decay amplitude as a sum of relativistic Breit-Wigner
propagators multiplied by form factors plus a term describing the angular
distribution of the two body decay of each intermediate state of a given spin.
Many amplitude analyses require detailed knowledge of the light-meson sector.
 In the case of a narrow, isolated
resonance, there is a close connection between the position of the pole on the
unphysical sheet and the peak we observe in experiments at real values of the
energy. However, when a resonance is broad and overlaps with other resonances,
this connection is lost.  The Breit-Wigner parameters measured on the real
axis (mass and width) can be connected to the pole-positions in the complex
energy plane only through models of analytic continuation.

A formalism for studying overlapping and many channel resonances has been
proposed long ago and is based on the  \emph{K-matrix} \cite{wigner,chung}
parametrization.
This formalism, originating in the context of two-body
scattering, can be generalized to cover the case of production of resonances in
more complex reactions \cite{aitch}, with the assumption that the two-body system in the
final state is an isolated one and that the two particles do not
simultaneously interact with the rest of the final state in the production
process \cite{chung}. The  \emph{K-matrix} approach allows us
to include the positions of the poles in the complex plane directly in our
analysis, thus directly incorporating the results from spectroscopy experiments. 
\par
\begin{figure}[htb]
  \begin{center}
  \includegraphics[width=0.47\textwidth]{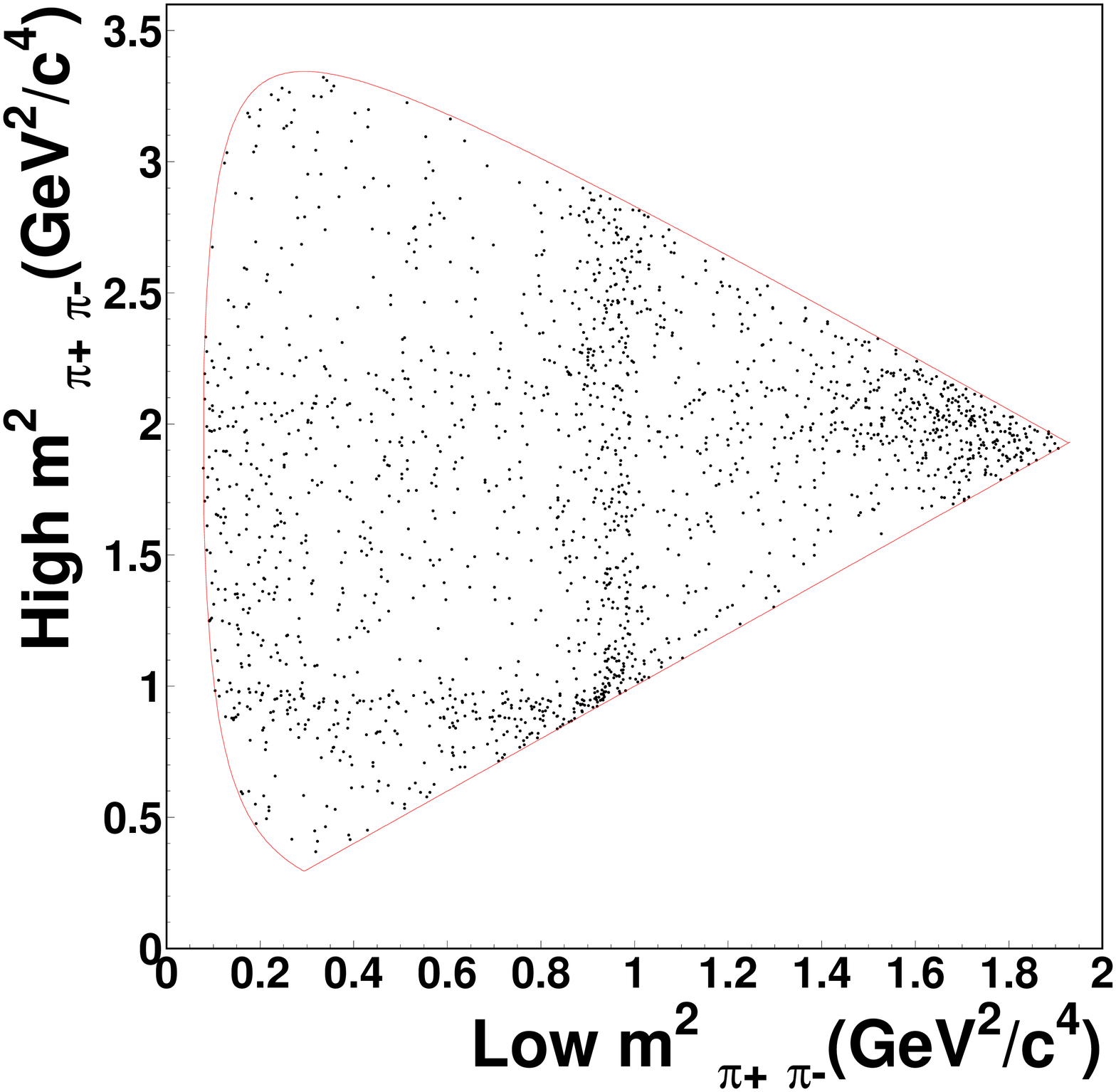}
  \includegraphics[width=0.47\textwidth]{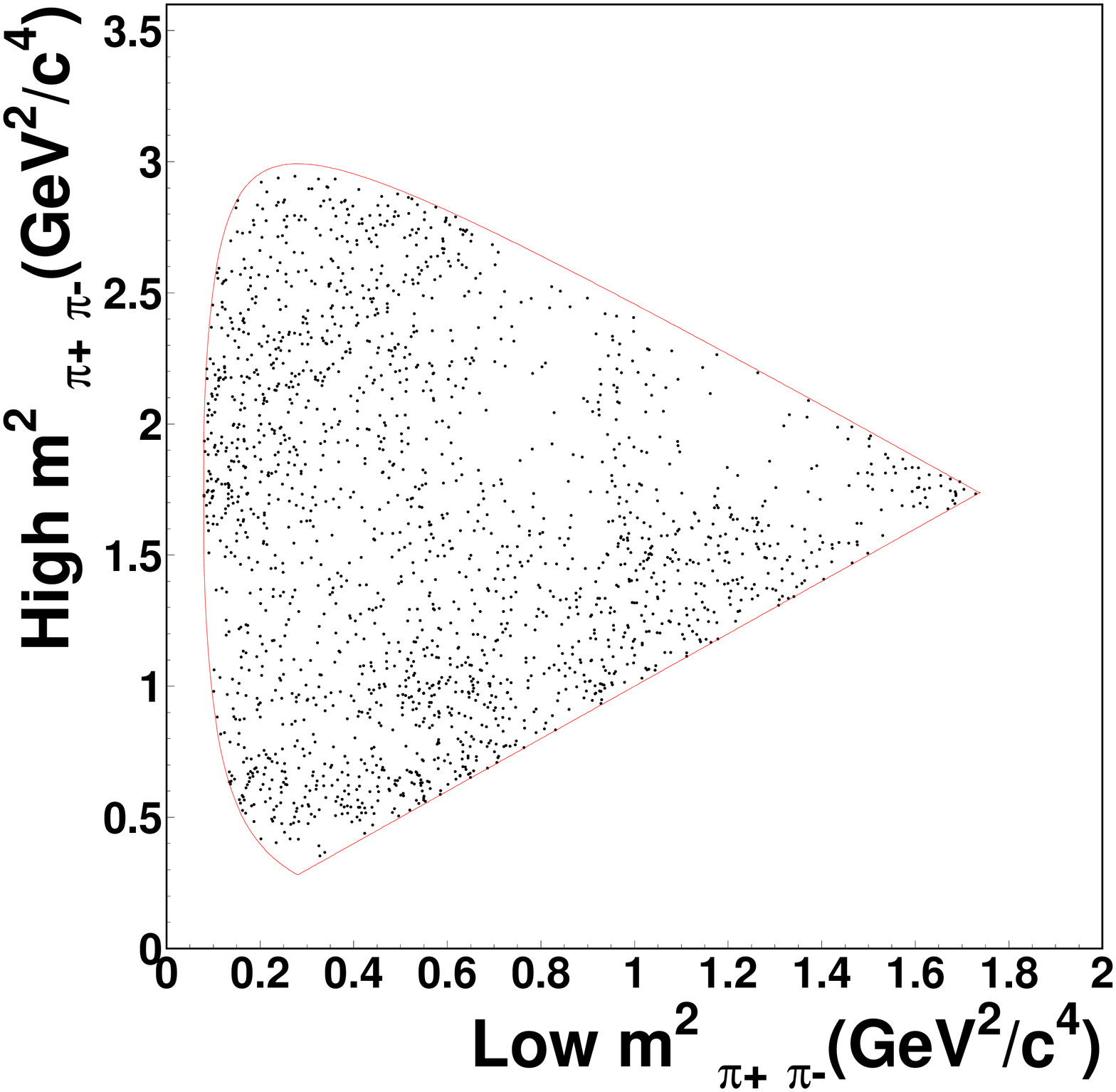}
  \end{center}
  \caption{\it  $D_s^+$ (left) and $D^+$ (right) Dalitz plots.}
  \label{dalitz}
\end{figure}
\begin{figure}
 \begin{center}
  \includegraphics[width=2.0in]{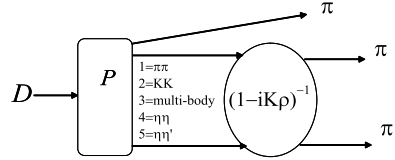}
  \caption{\it \emph{K-matrix} picture of a $D$ meson decay to 
   three pions, with
  a dipion in a isosinglet $S$-wave. 
 \label{kmCartoon}}
 \end{center}
\end{figure}

 Full details on event selection and analysis cuts are reported
in \cite{Link:2003gb}.
The Dalitz
plot analyses are performed on events within $2\,\sigma$ the nominal $D_s^+$ or
$D^+$ mass (Fig.~\ref{dalitz}).
The decay amplitude of the $D$ meson into the three-pion final state is written
as
$  A(D) = a_0 e^{i\delta_0} + F_1 + \sum_{i} a_i e^{i\delta_i} B(abc|r_i)$
where the first term represents the direct non-resonant three-body amplitude
contribution, $F_1$ is the contribution of  $S$-wave states and the sum is over
the contributions from the intermediate two-body non-scalar resonances.
$B(abc|r_i)$ are Breit-Wigner terms. 
The amplitude for the particular channel
$(00)^{++}_l\pi$ can be written in the
 context of the \emph{K-matrix} formalism as
$  F_l = (I-iK\rho)_{lj}^{-1}P_j$
where $I$ is the identity matrix, $K$ is the \emph{K-matrix} describing the
isoscalar $S$-wave scattering process, $\rho$ is the phase-space matrix for the
five channels, and $P$ is the ``initial" production vector into the five
channels. In this picture, the production process can be viewed (Figure
\ref{kmCartoon}) as consisting
of an initial preparation of several states, which are then propagated by the
$(I-iK\rho)^{-1}$ term into the final one. Only the $F_1$ amplitude is present
in the isosinglet $S$-wave term since we are describing the dipion channel.

We use the  \emph{K-matrix} parametrization of $(00)^{++}$-wave
scattering following  obtained through a global
fit of the available scattering data from $\pi\pi$ threshold up to $1900\,$MeV,
see \cite{anisar1}.
 The results
 are presented in Table~\ref{table_sol}.
\begin{table}[htb]
  \begin{center}
  \scriptsize
  \caption{\it Results on $D_s^+$ and $D^+\to\pi^+\pi^-\pi^+$  fit
  fractions and phases. Beside the first reported error, which is statistical,
  two systematic errors are quoted. The first one is from the measurement
  systematics and the second one is due to the particular solution chosen for the
  \emph{K-matrix} poles and backgrounds.}
  \label{table_sol}
\begin{tabular}{ccc}
\hline
 \multicolumn{3}{c}{$D_s^+$} \\
\hline
decay channel & fit fraction (\%) & phase (deg)\\
($S$-wave)\,$\pi^+$  & $87.04 \pm 5.60 \pm 4.17 \pm 1.34$ & 0 (fixed)\\
 $f_2(1270)\,\pi^+$  &  $9.74 \pm 4.49 \pm 2.63 \pm 1.32$ & $168.0 \pm 18.7 \pm2.5 \pm 21.7$ \\
 $\rho^0(1450)\pi^+$   &  $6.56 \pm 3.43 \pm 3.31 \pm 2.90$ & $234.9 \pm
 19.5 \pm 13.3 \pm 24.9$ \\
 \hline
 \multicolumn{3}{c}{$D^+$} \\
 \hline
decay channel & fit fraction (\%) & phase (deg)\\
($S$-wave)\,$\pi^+$    & $56.00 \pm 3.24 \pm 2.08 \pm 0.50$ & 0 (fixed)\\
 $f_2(1270)\,\pi^+$    & $11.74 \pm 1.90 \pm 0.23 \pm 0.18$   & $-47.5\pm 18.7 \pm
  11.7 \pm 5.3 $\\
 $\rho^0(770)\pi^+$   &    $30.82 \pm 3.14 \pm 2.29 \pm 0.17$   &$-139.4 \pm 16.5
 \pm 9.9 \pm 5.0$ \\
 \hline
 \end{tabular}
 \end{center}
\end{table}

\par
In conclusion,
the \emph{K-matrix} formalism has been applied for the first time to the charm
sector in our Dalitz plot analyses of the $D_s^+$ and $D^+\to\pi^+\pi^-\pi^+$
final states. Furthermore, the same
model is able to reproduce features of the $D^+\to\pi^+\pi^-\pi^+$ Dalitz plot
that otherwise would require an \emph{ad hoc} $\sigma$ resonance. In addition,
the non-resonant component of each decay seems to be described by known
two-body $S$-wave dynamics without the need to include constant amplitude
contributions.

The \emph{K-matrix} treatment of the $S$-wave component of the decay amplitude
allows for a direct interpretation of the decay mechanism in terms of the five
virtual channels considered:  $\pi\pi$, $K\bar K$, $\eta\eta$, $\eta\eta '$ and
$4\,\pi$. 
The resulting picture, for both $D_s^+$ and $D^+$ decay, is that the $S$-wave
decay is dominated by an initial production of $\eta\eta$, $\eta\eta'$ and
$K\bar K$ states. Dipion production is always much smaller. This suggests that
in both cases the $S$-wave decay amplitude primarily arises from a $s\bar s$
contribution such as that produced by the Cabibbo favoured weak diagram for the
$D_s^+$ and one of the two possible singly Cabibbo suppressed diagrams for the
$D^+$. For the $D^+$, the $s\bar s$ contribution competes with a $d\bar d$
contribution. That the $f_0(980)$ appears as a peak in the $\pi\pi$ mass
distribution in $D^+$ decay, as it does in $D_s^+$ decay, shows that for the
$S$-wave component the $s{\bar s}$ contribution dominates.
Comparing the relative $S$-wave fit fractions that we observe for $D_s^+$ and
$D^+$ reinforces this picture. The  $S$-wave decay fraction for the $D_s^+$
(87\,\%) is larger than that for the $D^+$ (56\,\%). Rather than coupling to an
$S$-wave dipion, the $d\bar d$ piece prefers to couple to a vector state like
$\rho^0(770)$ which accounts for $\sim30$\% of the $D^+$ decay.
This interpretation also bears on the role of the annihilation diagram in the
$D_s^+\to\pi^+\pi^-\pi^+$ decay. Our data suggest
that the $S$-wave annihilation contribution is negligible over much of the
dipion mass spectrum. It might be interesting to search for annihilation
contributions in higher spin channels, such as $\rho^0(1450)\pi$ and
$f_2(1270)\pi$.

\section{New measurements of the \phimndk{} form factor ratios }
\begin{figure}[t]
 \begin{center} 
 \includegraphics[width=3.in]{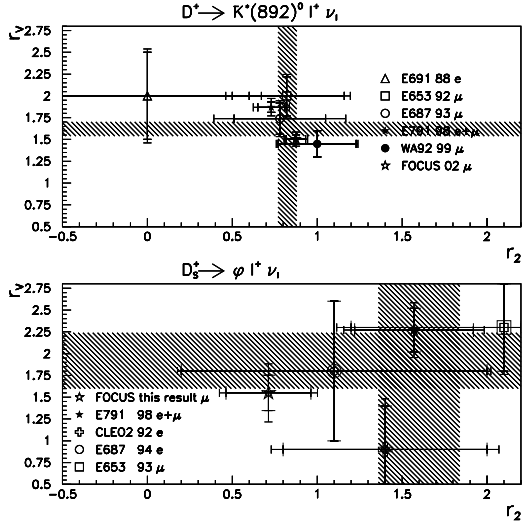}
\caption{\it Form-factor ratios comparison in previous data, and the new
 FOCUS result. World averages (not including this result) are also shown (shaded
 bands).
\label{ffratio3}}
\end{center}
\end{figure}

\begin{figure}
 \begin{center}
\includegraphics[width=2.in]{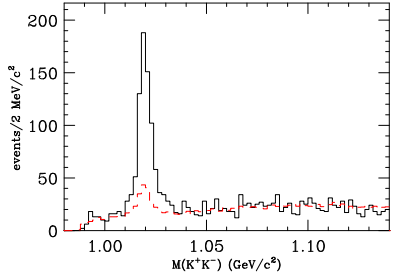}
\caption{ \it
The data is the solid histogram and $\ccb$ background Monte Carlo is
the dashed histogram. The $\ccb$ background Monte Carlo is 
normalized to the same number
of events in the sideband region
 1.04~\gevcsq{} $<$ \mkk{} $<$ 1.14~\gevcsq. 
\label{signal}}
\end{center}
\end{figure}

 The \phimndk{} decay amplitude
is described by four form factors with an assumed (pole form) \qsq{}
dependence.  
The \phimndk{} amplitude is then described by ratios of form factors taken
at \qsq{} = 0. The traditional set is: \rtwo{}, \rthree{}, and \rvee{}.
According to flavor SU(3) symmetry, one expects that the 
form factor ratios describing \phimndk{} should 
be similar to those describing \krzmndk{} 
since the only difference is an $\bar{s}$
spectator quark instead of a $\bar{d}$ spectator quark.  
The existing lattice gauge calculations \cite{theory} predict that the form
factor ratios describing \philndk{} should lie within 10\% of those describing
\krzlndk{}.    
Although the measured \rvee{} form factors are quite consistent between
\philndk{} and \krzlndk{}, 
there is presently a 3.3 $\sigma$ discrepancy 
between the \rtwo{} values measured
for these two processes with 
the previously measured \philndk{} value being a factor of about 1.8 times
larger than the \rtwo{} value measured for \krzlndk{} (
Figure \ref{ffratio3}). For a review and references see \cite{Bianco:2003vb}, 
while full details on event selection are found in \cite{Link:2004qt}.
The \mkk{} distribution for the \kkmndk{} candidates
is shown in Figure \ref{signal}.
\par

\par
The \rvee{} and \rtwo{} form factors were fit to the probability density 
function described by
four  kinematic variables 
(\qsq{}, \costhv{}, \costhl{}, and $\chi$) for decays
in the mass range $1.010 < \mkk{} < 1.030$.
We find \rvee{}=$\rvvalue{}$, \rtwo{}$=$\rtwovalue{}.
Our measured \rvee{} and \rtwo{} values for \phimndk{} are very consistent with
our measured \rvee{} and \rtwo{} values for \krzmndk{} \cite{focus_ff}. The
measurements reported here call into question the apparent inconsistency between
\rtwo{} values the \krzlndk{} and \philndk{} form factors present in previously
published data and are consistent with the theoretical expectation that the form
factors for the two processes should be very similar.

\section{L=1 excited charm meson spectroscopy}
High-statistics datasets from fixed-target and
 $e^+e^-$ colliders have recently provided the physics community with 
 a wealth of data on excited charm meson spectroscopy.
I report in this paper on new
results\cite{Link:2003bd} on L=1 $c\bar u, c\bar d, c\bar s$
states, pointing the reader to detailed reviews for an account  of the
experimental scenario \cite{Bartelt:1995rq,Bianco:2003vb}.
In the limit of infinitely heavy quark mass, 
 the heavy-light meson behaves analogously to the hydrogen atom, {\it i.e.,}
 the heavier quark does not contribute to the
 orbital degrees of freedom (which are completely defined by the light
 quark). 
 The angular momentum of the heavy quark is
 described by its spin $S_{Q}$, and  that of the light degrees of
freedom are described by  ${\bf j_{q}}\ = {\bf s_{q}}\, + \, {\bf L}$, 
where ${\bf s_q}$ is the light quark spin and ${\bf L}$ is the orbital
angular momentum of the light quark.
 The
quantum numbers ${\bf S_{Q}}$ and ${\bf j_{q}}$ 
are individually conserved. The quantum
numbers of the excited $L=1$ states are formed by combining 
${\bf S_Q}$ and ${\bf j_q}$. For $L=1$ we have $j_q=1/2$ and
    $j_q=3/2$. When combined with  ${\bf S_Q}$ they provide  two
    $j_q=1/2$ (J=0,1 where J is the total angular momentum of the
    excited charm meson) states, and two $j_q=3/2$ (J=1,2) states. 
    In this paper these
    four states will be denoted by 
    $D_0^*$, $D_1(j_q=1/2)$,
    $D_1(j_q=3/2)$ and $D_2^*$. 
    \par
\begin{figure}[t]
   \begin{center}
     {\includegraphics[scale=0.28]{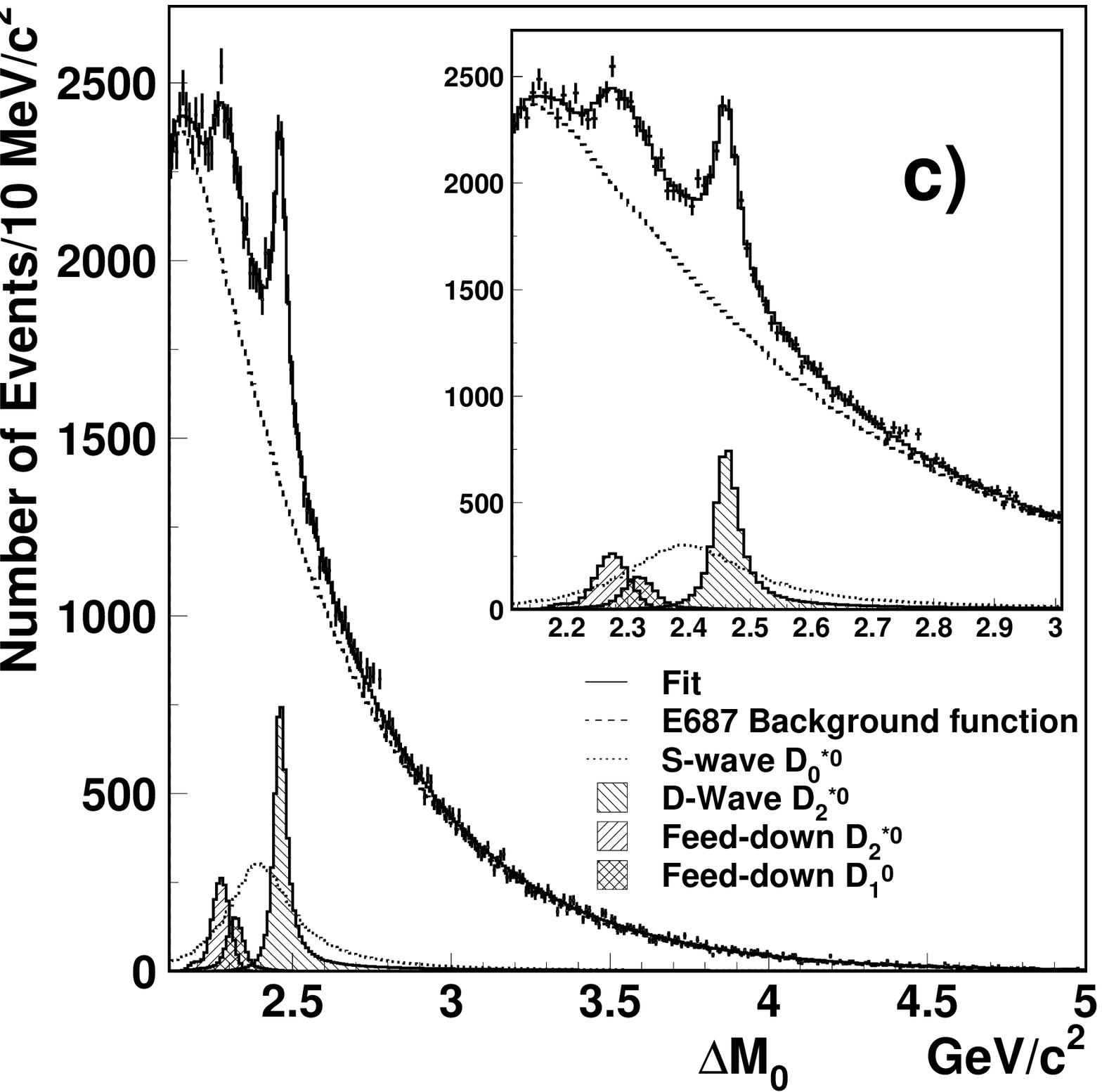}}
     {\includegraphics[scale=0.28]{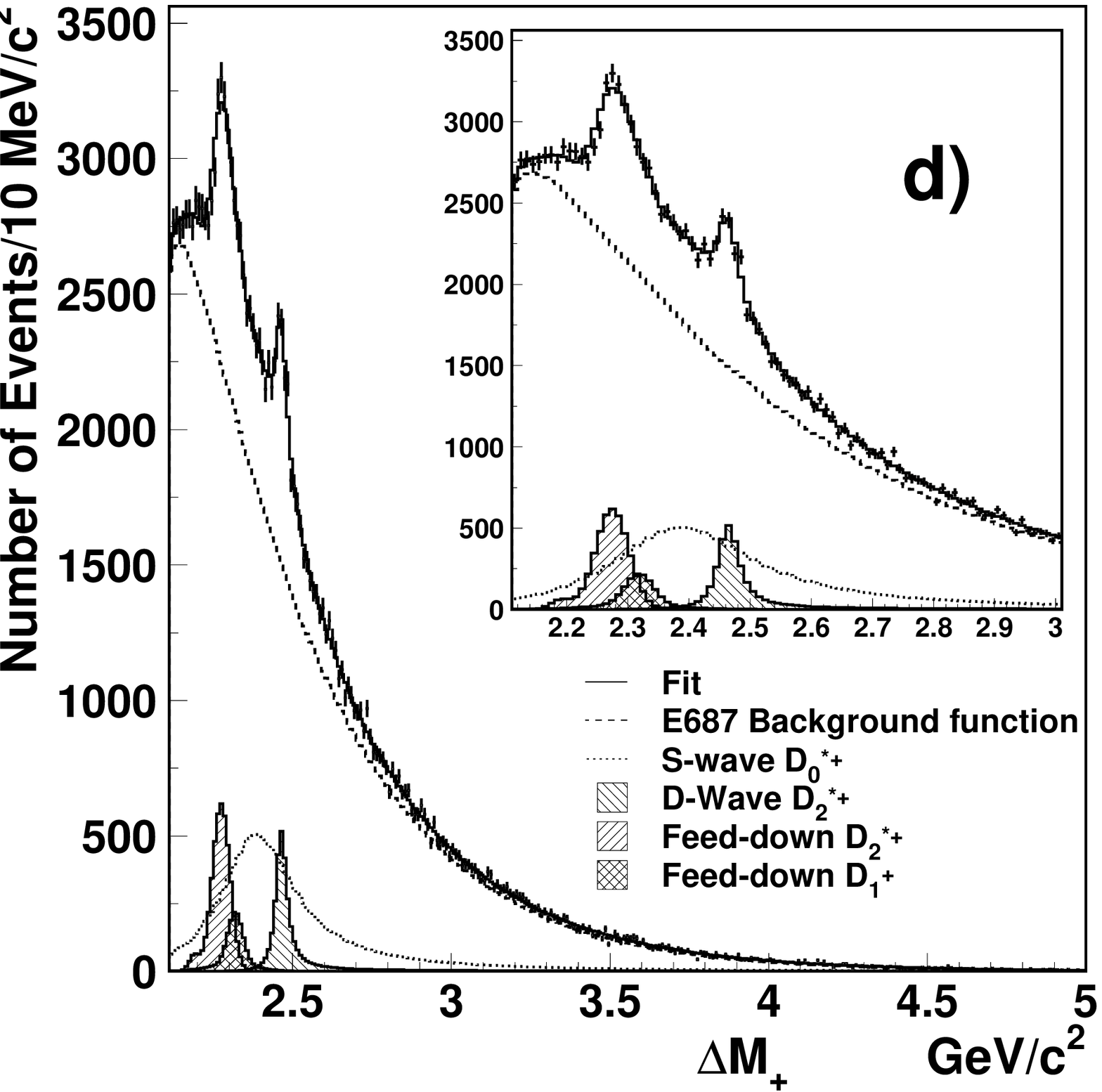}}          
     \caption{\it
     The fit to the $D^+\pi^-$ (left) and $D^0\pi^+$ (right) mass spectra including a term
 for an S-wave resonance.
       \label{fig:d0+-fin}}
   \end{center}
 \end{figure}

Analysis procedures are explained in detail in \cite{Link:2003bd}.
The $L=1$ charm mesons were reconstructed via
 $D^+\pi^-$ and $D^0\pi^+$ combinations.  The
    $D^0$ decays were reconstructed in the channels 
     $D^0\rightarrow K^- \pi^+ $ and
     $D^0\rightarrow K^- \pi^+ \pi^+ \pi^-$. The
$D^+$ decays were reconstructed in the channel 
  $D^+\rightarrow K^- \pi^+\pi^+$.
Our starting samples for these decay
modes  are 210,000, 125,000 and 200,000 
events, respectively.   
Figure~\ref{fig:d0+-fin}c) shows the distribution of the invariant mass
 difference 
  $$\Delta M_0\equiv M((K^-\pi^+\pi^+)\pi^-) - M(K^-\pi^+\pi^+) + 
   M_{\mathrm{{PDG}}}(D^+)$$
 where $M_{\mathrm{{PDG}}}(D^+)$ is the world average $D^+$  mass  
  \cite{pdg2004}.
 Figure \ref{fig:d0+-fin}c) 
 shows a pronounced, narrow peak near a mass $M \approx 2460 \mev$,
   which is consistent with 
 the  $D_2^{*0}$ mass. 
  The
 additional enhancement at $M\approx 2300 \mev$ is consistent 
 with feed-downs from the states
 $D_1^0$ and $D_2^{*0}$ decaying to $D^{*+}\pi^-$ when 
 the $D^{*+}$ subsequently decays to a $D^+$ and undetected neutrals.
 \par
 The mass difference 
$$   \Delta M_+  \equiv  M((K^-\pi^+,K^-\pi^+\pi^-\pi^+) \pi^+) - 
                M(K^-\pi^+,K^-\pi^+\pi^-\pi^+)  
             + M_{\mathrm{PDG}}(D^0)$$
 spectrum (Figure \ref{fig:d0+-fin}d) shows similar structures to the
 $\Delta M_0$ spectrum. The prominent peak is consistent with 
 a $D_2^{*+}$ of mass  $M \approx 2460 \mev$. 
 The additional enhancement at $M\approx 2300 \mev$ is again consistent
 with feed-downs. 
 \par
  We fit the invariant mass difference histograms
 with terms for the $D_2^{*0},~D_2^{*+}$ peaks, 
 $D_1$ and $D^*_2$ feed-downs, combinatoric background and the possibility of a
 broad resonance.  The broad resonance is necessary to obtain a fit to the data of
 acceptable quality (Fig.\ref{fig:d0+-fin} c-d). Our final results are shown in
 Table~\ref{tab:mass}. Our mass measurement of the broad state is higher than a
 recent measurement by BELLE~\cite{Abe:2003zm}. 
 Our result on the broad state have stimulated a 
 series
  of theory studies,
 which  try to reconcile the experimental picture of excited non-strange,
   and strange charmed mesons.
 \par

 \begin{table}
 \caption{\it Measured masses and widths for narrow and broad 
     structures in $D^+\pi^-$ and
      $D^0\pi^+$ invariant mass spectra. The first error listed is statistical and the second
      is systematic. Units for the masses and widths are $\mev$. 
       \label{tab:mass}}
    \scriptsize
     \begin{tabular}{lcccc} \hline
        & $D_2^{*0}$              & 
        $D_2^{*+}$             &
        $D_{1/2}^{0}$         &
         $D_{1/2}^{+}$ \\ 
                        &
                               &

                        &
                       &
                        \\
       \hline
       Yield            & 
       $5776 \pm 869 \pm 696$  &
        $3474 \pm 670 \pm 656$      &
         $  9810 \pm 2657$  &
         $18754 \pm 2189$\\
       \hline
       Mass             &
        $2464.5 \pm 1.1 \pm 1.9$ &
         $2467.6 \pm 1.5 \pm 0.76$   &
          $2407 \pm  21 \pm 35$ &
           $2403\ \pm  14 \pm 35$\\
       PDG03          &
        $2458.9 \pm 2.0$         &
         $2459 \pm 4$                &
          &
           \\
       \hline
       Width            &
        $38.7 \pm 5.3 \pm 2.9$   &
         $34.1 \pm 6.5 \pm 4.2$      &
          $240 \pm 55 \pm 59 $ &
           $283 \pm 24 \pm 34 $\\
       PDG03                        &
        $23 \pm 5$                 &
         $25^{+8}_{-7}$           &
          &
          \\
       \hline
     \end{tabular}
    \vfill
 \end{table}
\section*{Acknowledgments}
We wish to acknowledge the assistance of the staffs of Fermi National
Accelerator Laboratory, the INFN of Italy, and the physics departments of the
collaborating institutions. This research was supported in part by the U.~S.
National Science Foundation, the U.~S. Department of Energy, the Italian
Istituto Nazionale di Fisica Nucleare and Ministero della Istruzione
Universit\`a e Ricerca, the Brazilian Conselho Nacional de Desenvolvimento
Cient\'{\i}fico e Tecnol\'ogico, CONACyT-M\'exico, and the Korea Research
Foundation of the Korean Ministry of Education.


\begin{thebibliography}{0}

\bibitem{Link:2003gb}
 J.~M.~Link {\it et al.}  [FOCUS Collaboration],
 Phys.\ Lett.\ B {\bf 585} (2004) 200.

\bibitem{Link:2004qt}
 J.~M.~Link {\it et al.}  [FOCUS Collaboration],
 Phys.\ Lett.\ B {\bf 586} (2004) 183.

\bibitem{Link:2003bd}
 J.~M.~Link {\it et al.}  [FOCUS Collaboration], 
 Phys.\ Lett.\ B {\bf 586} (2004) 11.

\bibitem{Bianco:2003vb}
  S.~Bianco {\it et al.},
  Riv.\ Nuovo Cim.\  {\bf 26N7-8} (2003) 1.


\bibitem{Malvezzi:2003jp}
 S.~Malvezzi,
 AIP Conf.\ Proc.\  {\bf 688} (2004) 276.

\bibitem{wigner}
E.~P.~Wigner, Phys.~Rev.~70 (1946) 15.

\bibitem{chung}
S.~U.~Chung \emph{et al.},
  Ann.~Physik~4 (1995) 404.

\bibitem{aitch}
I.~J.~R.~Aitchison, Nucl.~Phys. A189, (1972) 417.

\bibitem{anisar1}
V.~V.~Anisovich and A.~V.~Sarantsev,
  Eur.~Phys.~J. A16 (2003) 229.
  
  

\bibitem{theory}
C.~W.~Benard, A.~X. El-Khadra, and A.~Soni, Phys. Rev. D45 (1992) 869. 
V.~Lubicz, G.~Martinelli, M.~S.~McCarthy, and C.~T.~Sachrajda, Phys. Lett. B 274 (1992) 415.
 \bibitem{focus_ff}
J.~M.~Link {\it et al.}  [FOCUS Collaboration],
  Phys.\ Lett.\ B 544 (2002) 89. 


\bibitem{Bartelt:1995rq}
  J.~Bartelt and S.~Shukla,
  Ann.\ Rev.\ Nucl.\ Part.\ Sci.\  {\bf 45}, 133 (1995).
 
\bibitem{pdg2004}
 S. Eidelman et al., Phys. Lett. {\bf B592}, 1 (2004). 


\bibitem{Abe:2003zm}
 K.~Abe {\it et al.}  [Belle Coll.],
 Phys.\ Rev.\ D {\bf 69}, 112002 (2004)

\bibitem{excitedTheory}
 A.~Hayashigaki and K.~Terasaki,
 hep-ph/0411285;
 C.~Quigg,
 hep-ph/0411058;
 S.~Godfrey,
 hep-ph/0409236;
 S.~Ishida, M.~Ishida, T.~Maeda, M.~Oda and K.~Yamada,
 hep-ph/0408136;
 R.~Ferrandes,
 hep-ph/0407212;
 T.~Mehen and R.~P.~Springer,
 hep-ph/0407181;
 P.~Colangelo, F.~De Fazio and R.~Ferrandes,
 Mod.\ Phys.\ Lett.\ A {\bf 19}, 2083 (2004);
 D.~Becirevic, S.~Fajfer and S.~Prelovsek,
 Phys.\ Lett.\ B {\bf 599}, 55 (2004);
 D.~Becirevic {\it et al.},
 hep-lat/0406031;
 Y.~I.~Azimov and K.~Goeke,
 Eur.\ Phys.\ J.\ A {\bf 21}, 501 (2004).
\end{thebibliography}
\end{document}